\documentclass{mn2e}

\usepackage[dvips]{graphicx}

\begin{document}

\title[Nitrogen in DLAs]{ Nitrogen Abundances in 
DLA Systems: The Combined Effects of SNII and SNIa in a Hierarchical
Clustering Scenario}

\author[Tissera et al.]{Patricia B. Tissera, $^{1,2}$,
Diego G. Lambas,$^{1,3,4}$, Sofia A. Cora,$^{1,5,6}$, and
Mirta B. Mosconi, $^3$ \\
$^1$ Consejo Nacional de Investigaciones Cient\'{\i}ficas
y T\'ecnicas.\\
$^2$ Instituto de Astronom\'{\i}a.
y F\'{\i}sica del Espacio, Argentina.\\
$^3$ Observatorio Astron\'omico
de la Universidad Nacional de C\'ordoba,  Argentina.\\
$^4$ John Simon Guggenheim Fellow.\\
$^5$ Postdoctoral Fellow of Fundaci\'on Antorchas at Max-Planck Institute for
Astrophysics, Germany.\\
$^6$ Observatorio Astron\'omico de La Plata, Argentina.
}

\maketitle

\begin{abstract}
          
The combined enrichment
of Supernovae II and I 
in a hierarchical clustering
scenario
 could produce regions with low N content respect to $\alpha$-elements
consistent
with observed values measured in Damped Ly-$\alpha$ (DLAs).
We have studied the formation of DLAs in a hierarchical clustering
scenario under the hypothesis that the building blocks of
current field galaxies could be part of the structures mapped
by DLAs. In our models 
the effects
of the non-linear evolution of the structure (which
produces bursty star formation histories, gas infall, etc.) and 
the contributions of SNIa and SNII
are found to be responsible of producing these  N regions 
with respect to the $\alpha$-elements.
Although SNIa are not main production
sites for Si or O, because of the particular timing
between SNIa and SNII, their contributions can help to produce
clouds with such  abundances. 
Consistently, we found the simulated low nitrogen DLAs to
have sub-solar [Fe/H]. 
We show that low nitrogen DLAs have experienced important
star formation activity in the past with higher efficiency than normal DLAs.
Our chemical model suggests that  SNIa play a relevant role in the 
determination of the abundance pattern of DLA
and, that  the observed low nitrogen DLA frequency
could  be explained taking into account the time-delay of $\approx $ 0.5 Gyr introduced
by these supernova to release metals.

\end{abstract}


\section{Introduction}

The study of chemical properties of the structure at different redshifts
have provided important clues for the problem of galaxy formation 
and evolution (Contini et al. 2002). In particular, 
DLAs have been used to probe
 the abundances of the neutral
Hydrogen over a large range of redshift becoming a powerful tool
to study the enrichment of the interstellar medium.
Several elements have been detected belonging to the Fe-group and
the so-called $\alpha$-elements.  
Owing to the fact that Fe and the $\alpha$-elements are thought to
have originated in different events: supernovae (SN)
 Ia  and II, respectively, and
consequently, there may be a lag between  their ejection into the interstellar
medium (ISM), their relative
abundances as a function of redshift could test the star formation (SF)
history of galaxies.

Recently, there have been new results on Nitrogen abundances in DLAs
which have fostered the discussion on its possible  nucleosynthesis 
sites, principally owing to  a group of low N (LN) abundance DLAs which lie
in the region between secondary and primary production according
to certain models (e.g., Henry, Edmund \& Koppen 2000).
The lasted data are given by 
Pettini et al. (2002, hereafter Pet02) and Prochaska et al. (2002, 
hereafter Pro02) who also discussed different possible mechanisms to explain
them.
 
Previous works that studied the N abundances in HII regions and Galactic stars
have shown the need for primary and secondary production, since
the N abundance clearly increases at a  faster rate than those of Si or O
for high metallicity.
However, complications arise due to the fact that N can be produced
in both massive and intermediate mass stars (IMS). The former would release it
shortly after their birth while the latest evolve over a longer
period of time estimated by Henry et al. (2000) in 250 Myr. 
This last mechanism would introduce a lag in the N release which
combined with the SF history, might leave characteristic
patterns in the abundance distributions.
Pet02 has suggested that this delay could explain the LN-DLAs if
IMS are the main production sites for the primary N component. However
because of  the shortness of this time delay,  
the estimated number  turned  out to be smaller than the number
of already observed LN-DLAs.
These authors estimated one out of seven 
 in the range  $z=6$ and $<z=2.6>$
for a $\Lambda$-CDM cosmology ($h=0.65$),
 while the observed frequency is   
 four out of ten DLAs with $<z=2.6>$.

We profit from having  a chemical enrichment model
within a cosmological context that allows a detailed
 description of the formation
and evolution of the structure together with the metallicity properties
of the ISM and the stellar population (Mosconi et al. 2001, hereafter Mos01). 
By using this chemo-dynamical code, 
Tissera et al. (2001, hereafter Tis01) and Cora et al. (2002) have shown
that the progenitors of current normal galaxy population in the field
in hierarchical clustering scenarios could give origin 
DLA clouds. The [Fe/H], [Zn/H] and [$\alpha$/H]
abundances were found to be 
 in very good agreement with the observed values and show
the same level of evolution if the observational filter defined
by Boiss\'e et al. (1998) is applied ($F=$[Zn/H] $+ {\rm log  \ N_{\rm HI}}$ with
$18.8 < F< 21$).
Our models include chemical feedback from SNII and SNIa. In this version, 
we have not
included  metal production from intermediate stars except
for those binary systems which are taken as progenitors of SNIa.
Hence, the results discussed here would show the effects of
the enrichment by SNII and SNIa with rates set by the SF 
history of the galactic objects which are given by their 
evolution within a hierarchical clustering   scenario.

\section{Analysis}

We performed simulations of $5h^{-1}$ Mpc side box with initial
conditions  consistent with  Standard Cold Dark Matter models 
($\Omega=1$, $\Lambda =0 $, $\Omega_b =0.10$, $\sigma_8 = 0.67$ and
 $H=100 h^{-1} \ {\rm Mpc/km/s} $ with $h=0.5$). 
We used $64^3$ particles with 
M$_{\rm p}=1.4 \times 10^{8} \ h^{-1} \ {\rm M_{\odot}}$. The gravitational
softening used is $1.5h^{-1}$  kpc and the minimum smoothing length is $0.75 h^{-1}$  kpc.
The SF scheme is based on the Schmidt law with a SF
timescale proportional to the dynamical one (Tissera et al. 1997).
Our simulations include the chemical model developed by Mos01, 
which allow the description of the enrichment of the stars and
gas as the galactic objects are assembled. 
For massive stars (M$> 8 {\rm M_{\odot}}$), we adopted the yields of Woosley \& Weaver (1995, WW95) and
for SNIa those of Thielemann, Nomoto \& Hashimoto (1993) with a time-delay
of 0.5 Gyr.
The SF and SN parameters used in this paper are those that, on average,
reproduce the abundance properties of nearby galaxies (Mos01).

We identified galaxy-like objects at different $z$ at their virial radius
($\delta \rho /\rho \approx 200$). In order to diminish numerical resolution
problems, we only  analyze the substructure  resolved
with more than 2000 particles within the virial radius.
In total, we will study 66 galactic objects  distributed within 
$0.26 \leq z \leq 2.35$. We draw random line-of-sights (LOS) through these
structures from three different random observers situated at $z=0$,
  and estimated the H column densities along them, generating 
198 LOS with 
 N(HI)$ > 2\times 10^{20} {\rm atoms/cm^{2}}$.
 More details
on the simulations and methods can be found in Tis01.
The relative  [K/J] abundance of  elements K and J
 in the simulated DLAs are estimated by adding the corresponding
metal masses in the gaseous component
belonging to each substructure along the LOS. 


%


We estimated the [N/Si] and (N/O) for the simulated DLAs in order
to confront them with observations from Pro02 and Pet02, respectively.
 Fig.1 shows [N/Si] versus [Si/H] and (N/O) versus
12 + (O/H) for simulated DLAs.
As it can be seen from this figure,  
the  N abundances for the simulated DLAs  are within the 
observational range. These abundances have been the results
of  the contribution from  massive stars estimated by using WW95 yields
which are the only N production site taken into account in our models.
WW95 ejecta have a significant dependence of the 
N production on  metallicity. 
Our results show that N nucleosynthesis in massive stars can account
fairly well for the observed N abundances in DLAs with low O (or Si) content.
Even more, there are some simulated LN-DLAs that
have very low N abundances with respect to the $\alpha$-elements at abundance
levels in agreement with observations.
The solid lines denote the limits that we have  chosen to divide 
the samples into a low and high N-DLA ones. 
We will focus on the analysis of these LN-DLAs and their origin.

\begin{figure}
\includegraphics[width=84mm]{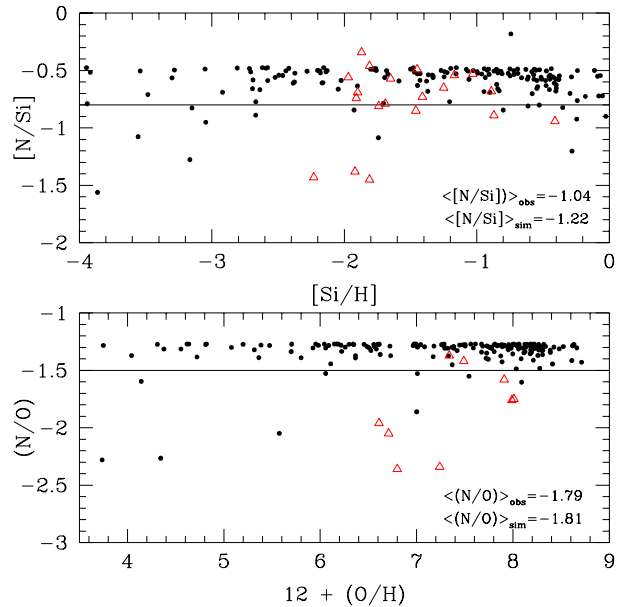}
\caption{ [N/Si] versus [Si/H] (upper panel) and
(N/O) versus 12 + (O/H) (lower panel) for the simulated
DLAs ( circles) and observations (triangles)
from Prochaska et al. (2002, upper panel) and
Pettini et al. (2002, lower panel).
The solid lines denote the chosen limits between
HN and LN-DLAs: [N/Si]$=-0.8$ dex and (N/O)$=-1.5$ dex. The mean abundances for the simulated and observed LN-DLAs
have been plotted.  }
\end{figure}

We estimated the  mass in stars along LOS, finding that LN-DLAs
have associated  stellar populations with masses
 larger than $10^{7} {\rm M_{\odot}}$
to up $\approx 3 \times 10^8{\rm M_{\odot}}$. Their impact parameters
are distributed within 5-20 kpc, consistently with those of DLAs 
with higher N content (HN-DLAs).
 From these two results we conclude that simulated
LN-DLAs  mapped neither very central regions nor regions with very low
 mass in stars.

For the purpose of further understanding how these low-N abundances  regions
arise in the simulations (which do not include  IMS enrichment), 
we analysed the Fe content respect to $\alpha$-elements such as Si.
The main  sites of Fe production  are SNIa which have a time-delay of
0.5 Gyr for its
release. Hence, Fe enrichment lags behind $\alpha$-elements  and is very much
linked to the SF and evolution histories of the systems.
In Fig.2 we have plotted [Si/H] versus [Fe/Si] for the simulated DLAs
distinguishing between those that have high/low [N/Si] (open/filled
circles) abundances. 
It can be clearly seen that the two groups segregate accordingly to
their $\alpha$-enhancement with LN-DLA showing sub-solar [Si/Fe]. 
This result could be explained by considering that 
W7 model for SNIa of Thielemann et al. (1993)
produces Si and O as well as Fe. Generally, the former are not taken into
account because their main sites of productions are SNII.
However, SNIa  contribute as well with $\alpha$-elements
 and it turns out that the combination
of SF history and the particular evolutionary
track of the simulated systems can make  the $\alpha$-element
 production of these events 
non-negligible respect to that of SNII, 
if the gas component is  mainly affected by SNIa.
This is clear from Fig.2 where the simulated LN-DLAs have sub-solar [Si/Fe]
demonstrating that Si lags behind Fe (that is mainly produced by SNIa), 
 while the rest of the simulated DLAs have the opposite behaviour with 
solar
or supra-solar abundances (showing that SNII production dominates in these
cases).
On the other hand,  LN-DLAs have  comparable [Si/H] abundances
 to those of the rest of the simulated  DLAs but,
have lower [N/Si], making evident a deficiency of N which is only produced by
SNII.

Hence, we found that these LN-HI regions may have been enriched
by a first generation of stars after which SF stops, and then,
by SNIa. Because the simulated structure forms in a 
hierarchical scenario which is consistently described by our models,  gas infall
can also play a role by being a continuous source of pristine material.
 As the result of the combination of these processes, 
the Si content is similar to those
of HN-DLAs because of the extra SNIa contribution, the Fe content
is high and the N is  low respect to the Si.
The fact that the simulated LN-DLAs tend to have a large fraction of baryons
in stars 
 and impact parameter 5 kpc $ < b <20$ kpc (with no trend to be in the outer regions) suggest 
that    they belong to regions that have depleted an important fraction
of their gas component. This may be the cause of the cease in the SF activity
needed to explain their abundance characteristics.

In order to further investigate this point, we estimated two  parameters along
the LOS: the DLA star formation efficiency ($\epsilon_{\rm DLA}$) defined
as the ratio between the masses in stars and in H  along the LOS, and the
mean formation redshift 
 of the stellar population ($z_{\rm DLA}$ given in redshift)
 associated with DLAs (i.e. measured along the LOS).
The assessment of the SF efficiency  was done in the following way.
We calculated the $\epsilon_{\rm DLA}$ that splits the LN-DLAs into 
a high and a low SF efficiency samples with equal number of members.
Then, we estimated the percentage of HN-DLAs which would belong to 
 the high $\epsilon_{\rm DLA}$ sample, finding that only 5 per cent have
such high efficiencies. 

For the mean formation redshfit,
 we found that stars in LN-DLAs have
$z_{\rm DLA}=4.29$ while those in HN-DLAs  have 
$z_{\rm DLA}=1.93$. Clearly, the stellar populations associated
with LN-DLAs are,  on average, significantly older.
The total  mass in stars associated to simulated LN-DLAs, 
their high redshift of formation and their high SF efficiency, 
 confirm  that these regions experienced important SF with high efficiency
 in the past, which afterward ceased.

We also found that LOS are not always good tracers of the global properties
of their host galaxies. For example, the range of mean (N/O)
for the simulated DLA galaxies  is $\approx [-1.5,-1.3]$, 
and we found no difference
between the mean SF efficiencies of DLA galaxies hosting LN-DLAs
and of  HN-DLAs.
 In the simulations, only the astrophysical properties of the matter along the LOS, correlates with
their abundance properties along LOS.

From Fig.1 we can also see that  observed DLAs with  low N have
slightly higher O and Si content than the corresponding simulated  
 LN-DLAs.
This can be understood because we are comparing DLAs at different
redshift intervals:  
[1.78,4.47] and [0.25,2.35]  for the observations and simulations, respectively.
 The simulated LN-DLAs
tend to be low HI density column regions. Higher N(HI)s in the
simulated redshift interval have had time to be
 more enriched moving up to higher N and Si abundances.

Note that the simulated N abundances have a much better
agreement with observations when compared to the Si than to the O abundances. 
We are not sure that this is an intrinsic problem of the nucleosynthesis of
the ejecta adopted or
if these elements are diferentially affected by dust, for example.
However, both Pet02 and Pro02 claim that differential 
dust depletion and obscuration
are not important for their analysis of the observed abundances.

\begin{figure}
\includegraphics[width=84mm]{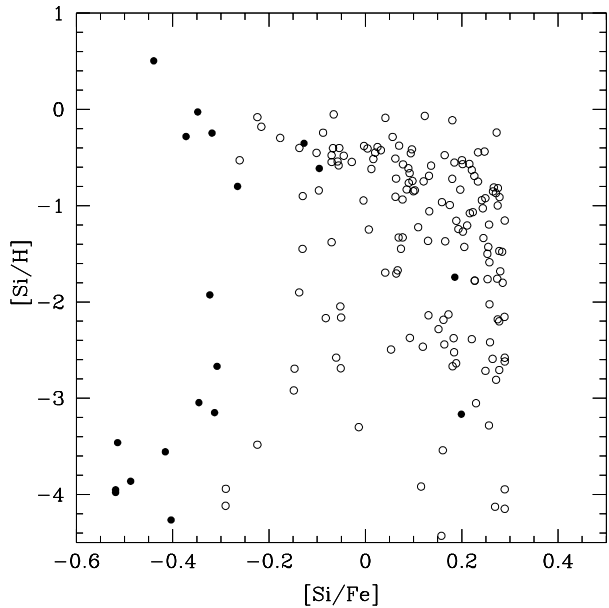}
\caption{
[Si/H] versus [Si/Fe] for the simulated DLAs,
distinguishing between those with high (open circles) and low (filled circles)
N content.
We use  [N/Si]$=-0.8$ as a  limit between 
these sub-samples.
}
\label{}
\end{figure}

\section{Discussion and Conclusions}

Our models describe the formation and evolution of galaxy-like objects
consistently with a cosmological model. In particular, we worked in
a hierarchical clustering scenario, hence, we followed the history
of assembly of galaxies including mergers and interactions.
As it has been shown (e.g., Tissera 2000),
 the SF history of galaxy-like objects in such scenarios
are significantly affected by their merger histories. 
Most objects show SF histories  with starbursts linked to mergers and interactions either
driven by secular evolution or their actual fusions (Scannapieco \& Tissera 2002).
Moreover, the relative mass distributions of baryons and dark matter
could be also modified  by mergers and, as a consequence, the response  to 
external factors such as interactions, could be different at different
stages of evolution affecting
their SF histories (Tissera et al. 2002).
Since the chemical history is directly linked to the SF process,
 the enrichment
of the ISM and SP requires the  adequate description of  these mechanisms. 
Another important process affecting the SF history
 and the metallicity of the systems
is gas infall which is  naturally taken them into account by our models.

We show that chemical enrichment by SNIa and SNII could
account for the N distribution in DLAs with $12 + {\rm (O/H)} \leq 8.5$,
including  low N DLAs \footnote{However, the  N production of SNII is not enough to generate
the observed steep  increase of this element respect to primary ones  in high
metallicity regions. We expect that the inclusion of IMS production will
produce the observed trend.}.  Our findings suggest  that LN-DLAs may
be the result of SNII and SNIa contributions together with
bursty SF  and gas infall.
These regions have high SF efficiencies and old SPs.
Simple calculations using the yields given by WW95 and Thielemann et al. (1993),
the Salpater Initial Mass Function and a ratio between SNI and SNII rates
of 3 (see Mos01) give that the Si production of SNII 
and SNI at low metallicity is of the same order.
We estimate that the contributions of two stellar populations with
mean metallicity of $Z=0.01 Z_{\odot}$ and $Z=0.0001 Z_{\odot}$
 could approximately  account for
the reported abundances.
(Note that these are rough values just to give an idea to
illustrate this fact). Our models keep  detailed 
records of the chemical content and production of the ISM and
the stellar populations as well as of mergers and gas infalls
 so that the abundances are obtained 
consistently with the history of formation and evolution.

There are two main achievements of this work. First, it clearly shows the
importance of using models that follows the non-linear evolution of the 
structure which affects the SF history and metal enrichment of both, stellar
and gaseous, components. 
Secondly, and as a consequence of the previous
fact, we suggest that SNIa may 
have played a more important role than hitherto thought. Following the reasoning
presented by Pet02, it can be estimated that
 the time delay for SNIa explosions ($\approx 0.5-1$  Gyr) could resolve
the inconsistency posed by the short life-time of IMS (250 Myr according to Henry et al. 2000). 
Note that the physical mechanism introducing the time-delay in our model
is different from that proposed by Pet02. In our case, SNIa are 
responsible of producing  lags among the elements.
Hence, a decrease in the SNIa life-time would lead to an increase
in the LN-DLA frequency, since these events would have had time
to take place in younger SPs as well. 
Following to Pet02's reasoning,  SPs younger than
0.5 Gyr would not have the chance to produce LN-DLAs. Hence,
in 1.8 Gyr (time between $z=2.6$ and $z=6$ in a $\Lambda$-CDM cosmology)
a period of 1.3 Gyr is left for SNIa to enrich the ISM. For a time-delay
of 0.5 Gyr, we estimate a LN-DLA frequency comparable to that deduced
by Pet02 from observations (i.e., approximately four out of ten).
We would like to stress that this SNIa time-delay also reproduces
the abundance patterns of galaxies at $z=0$  and the chemical evolution
and primary element contents of observed DLAs (Tis01).
Note that   a lower limit to SNIa life-time can be estimated
from this analysis, since
a shorter/longer time would produce too many/few LN-DLAs.
 
Finally, another important point derived from this analysis is
that the global properties of  DLA galaxies could be
being misinterpreted by looking at those of DLAs. LOS give information 
on very local regions which  might not trace the global
properties of the structure they are intercepting. 
A larger and careful statistical analysis  of DLAs is needed 
before drawing firm conclusions on the chemical
evolution of the Universe from the information provided by DLAs.


\section*{Acknowledgments}
We thanks the referee of this paper, Fabio Governato, for thoughtful
comments that helped to improved this paper.
This work was partially supported by 
CONICET, APCyT 
and   Fundaci\'on Antorchas.

\end{document}